# Curved-field optical coherence tomography: large-field imaging of human corneal cells and nerves

**Viacheslav Mazlin**[1,*], **Kristina Irsch**[2,3], **Michel Paques**[3], **Mathias Fink**[1] and **A. Claude Boccara**[1,**]

*[1]Langevin Institute, ESPCI Paris, PSL University, CNRS, 1 Rue Jussieu, 75005 Paris, France*

*[2]Vision Institute, Sorbonne University, CNRS, INSERM, 17 Rue Moreau, 75012 Paris, France*

*[3]Quinze-Vingts National Eye Hospital, 28 Rue de Charenton, 75012 Paris, France*

*\*mazlin.slava@gmail.com*

*\*\*claude.boccara@espci.fr*

**Abstract**

High-resolution optical imaging methods, such as confocal microscopy and full-field optical coherence tomography, capture flat optical sections of the sample. If the sample is curved, the optical field sections through several sample layers and the view of each layer is reduced. Here we present curved-field optical coherence tomography, capable of capturing optical sections of arbitrary curvature. We test the device on a challenging task of imaging the human cornea *in vivo* and achieve 10× larger viewing area comparing to the clinical state-of-the-art. This enables more precise cell and nerve counts, opening a path to improved diagnosis of corneal and general health conditions (e.g. diabetes). The method is non-contact, compact and works in a single-shot, making it readily available for use in optical research and clinical practice.

**One sentence summary**

Method for optical sectioning with arbitrary curvature is developed, enabling optical flattening of curved sample *in vivo*.

**Main text**

The cornea is the curved outermost part of the eye. Corneal transparency provides a unique opportunity to optically observe live microstructures of the eye and use them as indicators of ocular and general health. Clinical protocols rely on counts of cells and nerves, located in *en face* corneal planes and therefore require *en face* corneal images with a large field-of-view (FOV). Unfortunately, the FOV of existing high-resolution clinical modalities, such as *in vivo* confocal microscopy (IVCM) and specular microscopy (SM), is optically limited to 0.4 mm × 0.4 mm. Emerging *in vivo* research devices originating from the conventional Fourier-domain OCT such as UHR-OCT(*1*),(*2*), GDOCM(*3*), µOCT(*4*) can increase the FOV up to about 1 mm x 1 mm, however the cellular mosaics on that scale are free of motion artifacts only in anaesthetized animals, immobilized during the prolonged laser beam scanning in *en face* plane. The faster Fourier-domain full-field optical coherence tomography (FD-FF-OCT) modality can capture images free of scanning-related artifacts with 0.615 mm × 0.615 mm FOV, but requires costly hardware, such as a high-speed camera and swept-source laser(*5*). Alternative to the above backscattering detection methods, the retroillumination microscopy by Weber and Mertz(*6*) demonstrates a 0.820 mm × 0.580 mm FOV.

Recently, we developed an *en face* optical sectioning method, namely *in vivo* time-domain full-field optical coherence tomography (TD-FF-OCT)(*7*),(*8*), which can obtain cell-detail *in vivo* corneal images with 10× larger viewing area at 1.2 mm × 1.2 mm. Nevertheless, as the cornea exhibits natural curvature, the flat field optically sections through several corneal layers at once, keeping the FOV of each fine curved corneal layer limited. Here we demonstrate a method, termed curved-field optical coherence tomography (CF-OCT), which can capture optical sections of arbitrary curvature. Applied to the *in vivo* human cornea, this method enables full-field views of the curved sub-basal nerve plexus (SNP) and endothelial corneal layers at 1.13 mm × 1.13 mm and beyond. Moreover, high *en face* imaging speed of CF-OCT (0.6 billion pixels/second) ensures that images are free of eye or head movement artifacts. Large-field views of the SNP, obtained in a non-contact way, open a path for simple and precise monitoring the progression of diabetes, known to alter the corneal nerve density and tortuosity(*9*). In addition, larger views of the endothelial cell mosaic are expected to improve the outcome of corneal transplantation and cataract surgeries, which are today performed upon confirming endothelial health and a minimal cell count(*10*).

To achieve the curved optical sectioning, we implemented a simple optical lens in the conventional interferometric TD-FF-OCT design (Fig. 1). The light from the near-infrared (NIR) 850 nm light-emitting diode (LED) is first separated by the beam splitter (BS) into the sample and reference arms of the interferometer and then focused by the 10× air microscope objectives with moderate numerical aperture (0.3 NA) on the sample and on the front surface of the curved optical lens, acting as a curved mirror. The reflected light from this curved mirror and the backscattered light from the different layers of the sample recombine on the BS, producing interference. However, due to the low-temporal coherence of the LED, interference happens only for the backscattered light originating from the curved section of the sample, which matches with the curved mirror in terms of the path length to the BS. All the light (interfering and non-interfering) is detected by the 2D camera, composed of 1440 × 1440 pixels. Following the conventional TD-FF-OCT tomographic image retrieval scheme(*11*)·(*12*), the camera rapidly (at 550 frames/second, 1.75 ms per image) captures two consecutive images with different optical phases, modulated by the piezo element, and the simple algebraic post-processing on the images reveals the curved optical section of the sample. The above general approach can be used to obtain optical sections of any shape. We experimentally confirmed the 1.7 µm lateral and 7.7 µm axial resolutions of the optical system.

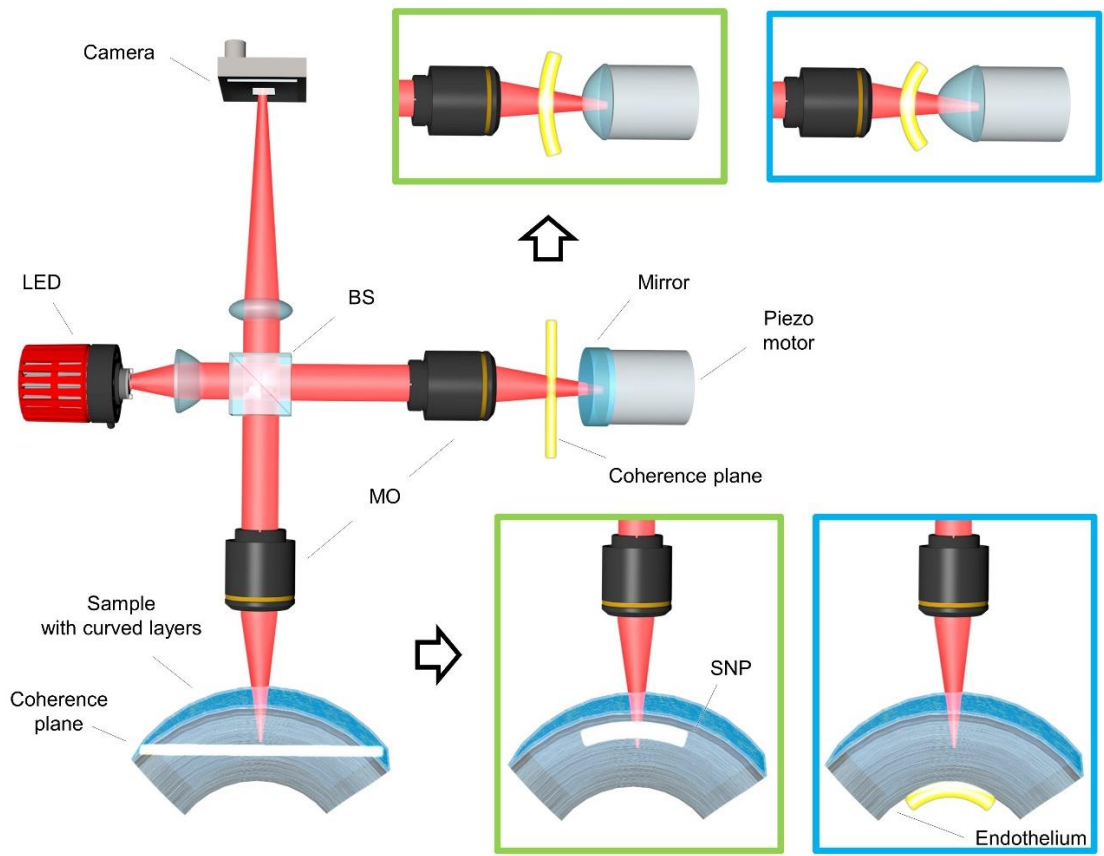

**Fig. 1. Comparison of curved-field OCT and conventional time-domain full-field OCT designs.** The optical interferometer is equipped with the incoherent LED light source, 2D camera and microscope objectives (MO). The location of the coherence plane, corresponding to the position of the reference arm is depicted in yellow. Use of a simple optical lens instead of a flat mirror allows to curve the coherence gate and obtain the optical sections of arbitrary curvature. Green embedding – CF-OCT configuration with a lens having 7.7 mm radius of curvature, optimal for optical flattening the 7.79 ± 0.27 (SD) mm curved anterior cornea. Blue embedding – CF-OCT

configuration with a lens having 6.2 mm radius of curvature, sufficient for optical flattening the 6.53 ± 0.25 (SD) mm curved posterior cornea.

In application to human corneal imaging, we use two different lenses with radii of curvature of 7.7 mm and 6.2 mm to optically flatten the anterior and posterior cornea, which exhibit the normal curvatures of 7.79 ± 0.27 (SD) mm and 6.53 ± 0.25 (SD) mm, respectively(*13*). We experimentally confirmed the close matching of mirror and corneal curvatures by looking at the density of the fringes, analogous to optical metrology using, for example, a Twyman Green interferometer. More precisely, the light beams, reflected from the surfaces of similar shape in the two interferometric arms, will have similar phase delay over the entire field. The resulting interference pattern on the camera will show only a few fringes, indicating an optical path delay of a few half-wavelengths. This is the case with the conventional TD-FF-OCT with identical flat reflectors in both interferometer arms (Fig. S1A) and CF-OCT with the curved reflector matching to the shape of the eye up to several μm, which is confirmed by counting about 10 fringes across the field from the center of the curvature (Fig. S1D,F). On the contrary, when the surface curvatures in both arms do not match, the fringe frequency that is low at the curvature apex is gradually increasing to the edges of the field, reflecting an increase in optical path length difference between the arms (Fig. S1B,C,E).

To achieve a correct curved optical sectioning, the apex of the curved mirror should match to the corneal apex, however the latter is constantly shifting due to *in vivo* ocular and head movements. In order to match the apexes, we mounted the interferometer on a 3-axis motorized stage, controlled with a joystick, while using the images as feedback (Fig. S2). More precisely, when imaging the SNP and endothelium, we looked at the mirror-like fringe patterns at the outermost and innermost corneal surfaces, respectively. The high fringe density along the X or Y axes indicated misalignment, which could be compensated for, by shifting the interferometer along the corresponding direction, until the fringe density decreased.

The proof of concept test was carried out on a healthy subject (female, aged 37 years), which was confirmed by routine eye examination in the hospital preceding the experiment. Approval for the study was obtained (study number 2019-A00942-55), in conformity with French regulations, from the CPP (Comité de Protection de Personnes) Sud-Est III de Bron and ANSM (Agence Nationale de Sécurité du Médicament et des Produits de Santé). Prior to experimental procedures, which adhered to the tenets of the Declaration of Helsinki, informed consent was obtained from the subject after the nature of the study were explained. During the experiment, the subject was asked to rest the chin and temples on a standard headrest, while looking at a fixation target. Examination was non-contact and without prior introduction of any cycloplegic or mydriatric agents, nor topical anesthetics. The pulsed light irradiance was below the maximum permissible exposure (MPE) levels of up-to-date ISO 15004-2:2007 (18% of MPE for cornea and 1% of MPE for retina) and ANSI Z80.36-2016 (1.3% of MPE for cornea and 1% of MPE for retina) (for more details see Methods). Illumination was comfortable for viewing, due to the low sensitivity of the retina to NIR light.

Figure 2 illustrates the comparison between IVCM (clinical state-of-the-art), TD-FF-OCT and CF-OCT images of the SNP, acquired from the same heathy subject. CF-OCT revealed the 2 – 4 μm thick corneal nerves within 1.13 mm × 1.13 mm FOV, which is about 10× larger (in area) than the clinical IVCM. Although this field is smaller comparing to the recent research version of IVCM that was able to achieve enlarged FOV of more than 2 mm × 2 mm by using a moving fixation target and mosaicking of more than 1000 conventional IVCM images (*14*)·(*15*), the latter approach required perfect fixation of the subject during a prolonged period (about a minute), which is impossible in many clinical cases. As another limitation, IVCM requires introduction of ocular anesthetic and physical contact with the patient's eye, resulting in discomfort for the patient, increased risk of corneal damage and appearance of corneal applanation artifacts in the images. On the contrary, CF-OCT is a non-contact modality, and the increased FOV, which is rapidly captured in 3.5 ms, simplifies locating of the same region of interest over time and increases the accuracy of measured nerve density, which is an important disease indicator, such as for diabetes progression.

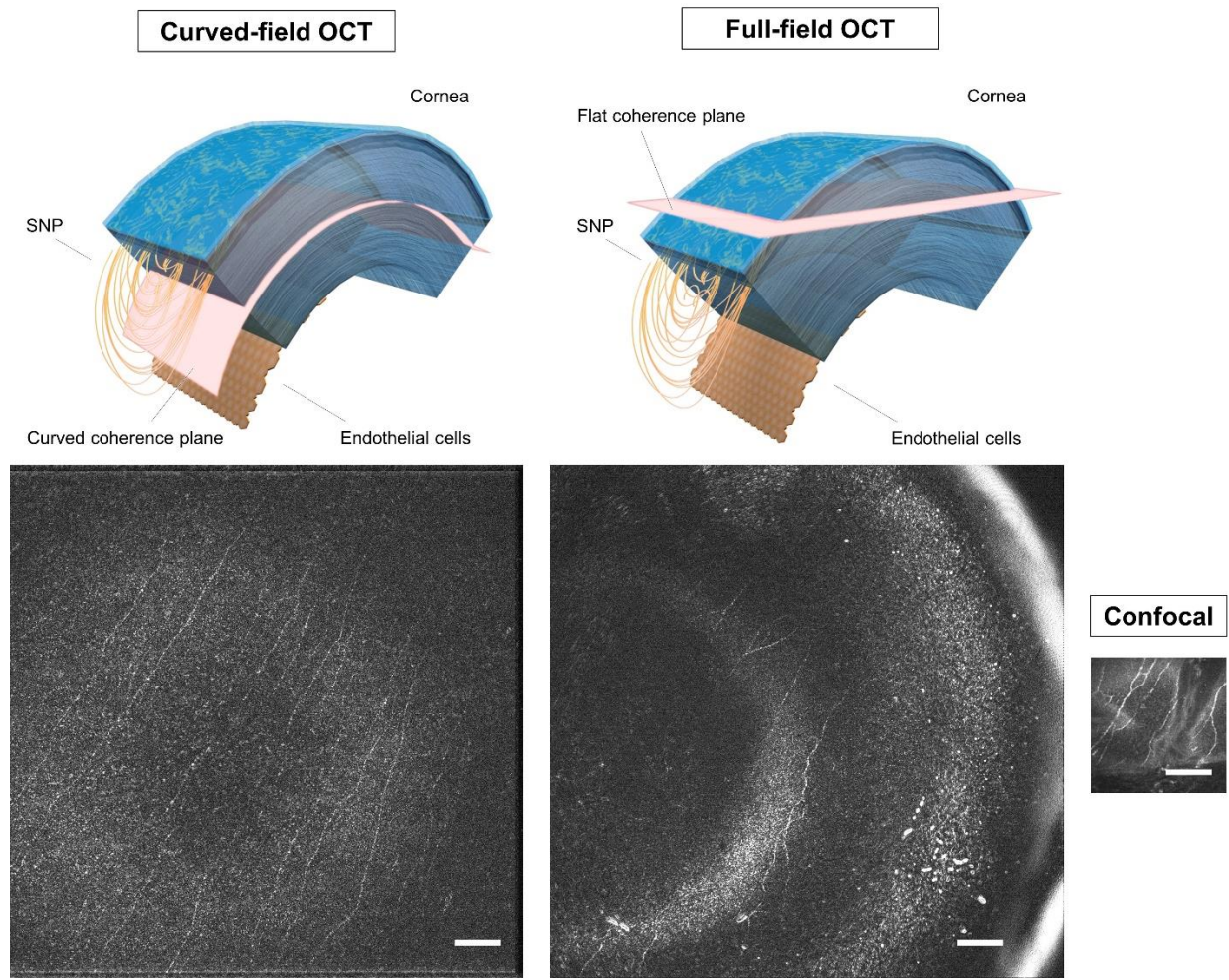

**Fig. 2. Curved-field OCT versus full-field OCT and confocal microscopy for imaging of the SNP in the human cornea *in vivo*.** By matching the curvature of optical sectioning with that of the cornea, CF-OCT substantially increases the FOV of the SNP layer, in comparison to the state-of-the-art TD-FF-OCT and CM. Non-contact CF-OCT is free from corneal applanation artifacts, which complicate SNP imaging in contact CM. Images were obtained from the same subject. All scale bars are 0.1 mm.

Figure 3 shows the comparison of endothelial images acquired with clinical specular microscopy (SM), TD-FF-OCT and CF-OCT. While SM and TD-FF-OCT are both non-contact methods and show the increased FOV over CM, CF-OCT demonstrates further improvement, reaching a 1.13 mm × 1.13 mm view of the endothelium. We performed quantitative comparison. We counted 1743 cells over the best focus 0.85 mm circular area within CF-OCT field, equivalent to a cell density of 3072 cells/mm$^2$, in agreement with the literature for normal human corneas(*16*). Similarly, with SM, we counted 387 cells and measured the matching cell density of 3096 cells/mm$^2$. Note that defocus decreases the sharpness of cells at the edge of CF-OCT FOV, despite the fact that the lateral resolution of TD-FF-OCT with spatially incoherent illumination is less affected by optical aberrations(*17*). The reason is that defocus is present in both arms of the CF-OCT interferometer, therefore optical aberrations affect CF-OCT as much as a conventional microscope.

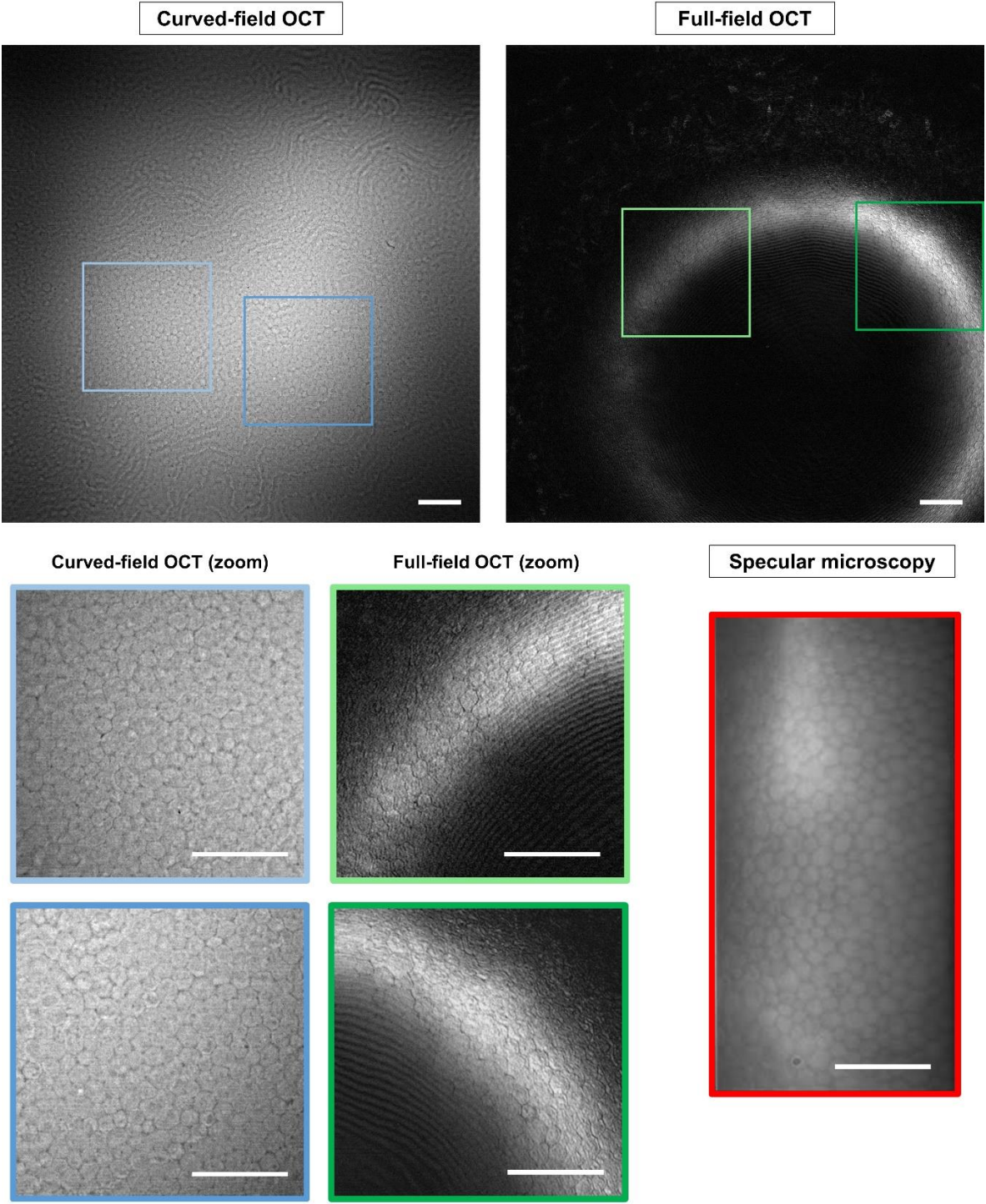

**Fig. 3. Curved-field OCT versus full-field OCT and specular microscopy for imaging of the endothelium in the human cornea *in vivo*.** By matching the curvature of optical sectioning with the curvature of the cornea, CF-

OCT substantially increases the FOV of the corneal endothelial layer, in comparison to the state-of-the-art TD-FF-OCT and SM. All scale bars are 0.1 mm.

Further note that obtaining endothelial images in Figure 3 required an additional post-processing step, highlighted in Figure S3. More precisely, the endothelium, being the last corneal layer, acts as a mirror, producing strong regular interference fringes in the camera and tomographic images. We remove these fringes by first averaging about 15 tomographic images to reduce the sharpness (and therefore the spatial frequency) of the fringe borders, and then by applying a disk mask filter in the Fourier domain.

Next we explored the following two possibilities for simplified use of the device in clinical setting: 1) single frame imaging without averaging, 2) using a single curved lens for both anterior and posterior corneal imaging.

Figure 4 shows an image of the endothelium, obtained from a single tomographic frame, captured in 3.5 ms. As image averaging was not performed, the Fourier mask filter needed to be extended to filter the higher spatial frequencies, comparing to the filter in Figure S3, to remove the sharp fringe borders. This filtering affected the image contrast, nevertheless, the cell count could be performed similarly, as in the averaged image.

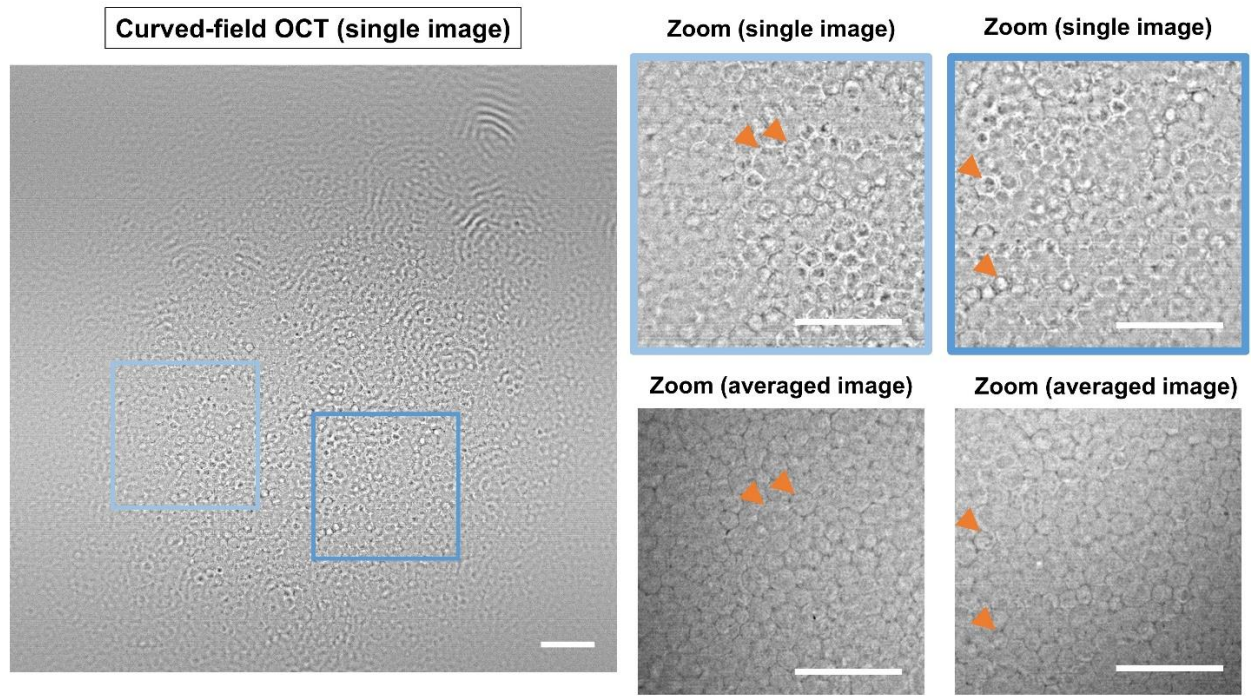

**Fig. 4. Comparison of curved-field OCT images captured in a single shot (3.5 ms) and averaged (52.5 ms) from *in vivo* human cornea.** Single shot image of the endothelium, obtained by subtracting the two camera images

and by Fourier filtering with a mask extended to higher spatial frequencies, has a different contrast comparing to SM or averaged CF-OCT images. Nevertheless, the same cells are revealed. Images were obtained from the same subject. All scale bars are 0.1 mm.

Figure S4 explores the possibility of using a single 7.7 mm lens for imaging the anterior and posterior cornea instead of the two separate lenses of 7.7 mm and 6.2 mm radii of curvature. Having a matching curvature with the anterior cornea, the 7.7 mm lens produces fringes with densities increasing to the edge of the field in the posterior cornea. The spatial frequency of the fringes at the edge is close to the frequency associated with cell borders, therefore the fringes cannot be filtered in the Fourier domain without affecting the visibility of cells. As discussed above, one can still remove the fringes by reducing the fringe border frequency through averaging of several 2-phase tomographic images before filtering in the Fourier domain. Alternatively, we show that in occasional moments, when the eye is static during the acquisition of 4 images (7 ms), the 4-phase modulation scheme can be used to substantially decrease the fringe sharpness comparing to the 2-phase scheme and, after filtering in the Fourier domain, reveal the cell mosaic.

Imaging the endothelium on a large scale has the potential to improve the outcome for patients undergoing corneal transplantation or cataract surgeries, which are today performed upon confirming endothelial health. Large FOV reduces the chance of missing the disease affected area and increases the precision of cell counts, improving diagnosis in a variety of ocular conditions, including Fuchs' dystrophy, endothelial trauma, iridocorneal endothelial syndrome, keratoconus, and others.

Future versions of the CF-OCT device can benefit from incorporating automatic cell-counting techniques(*18*), reducing the screening time in busy clinical settings. The in-focus FOV, presently limited to 0.85 mm, can be greatly extended by utilizing microscope objectives with smaller NA. For example, reduction in NA from 0.3 to 0.2 will increase the depth of focus by more than twice, while reduced lateral resolution from 1.7 μm to 2.5 μm will still be sufficient to resolve nerves and cells, as was confirmed in(*1*) and(*5*). Aside from ophthalmic imaging, CF-OCT may prove useful for non-contact exploration of various *in vivo* as well as *ex vivo* human and animal tissues exhibiting a curved structure.

**Acknowledgements**

We would like to acknowledge the advisory support of the Quinze-Vingts National Eye Hospital. We are also grateful to Cristina Georgeon, Marie Borderie, and Roxane Cuyaubère for assistance with acquiring the confocal and specular microscope images. We further thank Kate Grieve for valuable discussions. **Funding:** This work was supported by the HELMHOLTZ synergy grant funded by the European Research Council (ERC) (610110), a CNRS pre-maturation grant, Region Ile-De-France fund SESAME 4D-EYE [EX047007], French state fund CARNOT VOIR ET ENTENDRE [x16-CARN 0029-01] and French state fund IHU FOReSIGHT [ANR-18-IAHU-0001]. **Author contributions:** V.M., M.F. and A.C.B. conceptualized the general idea. V.M., K.I. and A.C.B conceived and developed the optical design. V.M. and K.I. performed the imaging experiments. V.M., K.I. and A.C.B. analyzed the acquired data. M.P. guided the clinical scientific part of the work and provided regulatory, financial and resources (facility, where the study was conducted) support. V.M. wrote the manuscript, and all the authors contributed with edits and revisions. **Competing interests:** Authors declare no competing interests. **Data and materials availability:** All data is available in the main text or the supplementary materials.

**Methods**

**Curved-field OCT device**

The curved-field OCT (CF-OCT) device is based on an interference microscope in a Linnik configuration with identical microscope objectives (LMPLN10XIR, Olympus, Japan) in the two arms of the interferometer. Objectives have 10× magnification, a numerical aperture (NA) of 0.3, and provide a high lateral resolution of 1.7 µm. This lateral resolution was estimated theoretically from the Rayleigh criterion and experimentally confirmed with a resolution target and by measuring the diameters (FWHM) of 80 nm gold nanoparticles imaged on a glass plate, located at the focal plane of the microscope objective. The working distance of the objectives was 18 mm, sufficient to avoid the risk of accidental physical contact with the eye. Illumination was provided by an NIR 850 nm light-emitting diode (LED) source (M850LP1, Thorlabs, USA). The axial resolution of 7.7 µm in the cornea was estimated from the experimentally measured spectral bandwidth of the LED (30 nm) with a spectrometer (CCS175/M, Thorlabs, USA) and by using the average corneal refractive index of 1.376. Light from the LED is collected by an aspheric condenser lens (ACL12708U-B, Thorlabs, USA) and is focused on the back focal plane of the objective. Before entering the objective, light from the source is equally separated by the 50:50 beam splitter (BS) (BS014, Thorlabs, USA) into the sample and reference arms of the interferometer. The objective in the reference arm focuses light onto the curved surface of the plano-convex lens with either 7.7 mm (LA1540, Thorlabs, USA) or 6.2 mm radius of curvature (LA1576, Thorlabs, USA), which plays the role of a single mirrored surface with 4% reflectivity. A low reflectivity value is chosen to achieve high detection sensitivity, which is maximized, when the total reflectivity of the sample and reflectivity of the reference mirror match, as indicated by signal-to-noise calculations of the time-domain full-field optical coherence tomography (TD-FF-OCT) method(*12*). Total reflectivity from all corneal layers, estimated from the Fresnel relations, is around 2% (*19*). By using a reference mirror with a reflectivity of 4%, we can expect sensitivity close to the ideal condition(*12*) at least for healthy or pathologic patients with a low scattering cornea. Position of lens is fine-tuned to have the curvature apex matching with the center of the camera FOV. Light in the reference arm, reflected from the curved lens surface, and light in the sample arm, backscattered from the different layers of the cornea, are collected by the objectives, and get recombined on the BS. This results in interference, but only for light coming from the curved lens surface and the light coming from the corneal layer, which match in terms of optical path length. The temporal coherence length of the light source determines the thickness of the interference fringe axial extension and therefore the optical sectioning precision, which is 7.7 µm. Interfering light and non-interfering light, arising from other planes of the cornea, are focused on a high full-well capacity (2Me-) 1440×1440 pixel CMOS camera (Q-2A750-CXP, Adimec, Netherlands) by a 250 mm tube lens (AC254-250-B, Thorlabs, USA). The sensor captures 2D images at a rate of 550 frames/second with an exposure time of 1.75 ms. In order to extract the interfered light from the background, we use a 2-phase (or 4-phase) shifting scheme, where we rapidly modulate the position of the reference mirror using a piezo mirror-shifter (STr-25/150/6, Piezomechanik GmbH, Germany), synchronized with the camera, to capture two (or four) consecutive images, which have a $\pi$ (or $\pi/2$) phase shift, and subtract them. The absolute value of the resulting image contains only interfering light from a 7.7 µm thick section in the cornea.

The interferometer is positioned on the two high-load lateral translation stages (NRT150/M, Thorlabs, USA), controlled by a driver (BSC202, Thorlabs, USA) with a joystick (MJC001, Thorlabs, USA). Beneath, a motorized high-load vertical translation stage (MLJ150/M, Thorlabs, USA) is used to position the entire device vertically. The reference arm of the CF-OCT device is mounted on a voice-coil translation stage (X-DMQ12P-DE52, Zaber, Canada), used for preliminary aligning to image the selected corneal layer of interest All the device control and processing take place on a single personal computer (PC).

**In vivo imaging**

Informed consent was obtained from the subject and the experimental procedures adhered to the tenets of the Declaration of Helsinki. Examination was non-contact and no topical anesthetics were introduced onto the eye. Light illumination, visible as a dim red circular background, was comfortable for viewing, due to the low sensitivity of the retina to NIR light. Light exposures measured 18% of maximum permissible exposure (MPE) for the cornea and only 1% of MPE for the retina (according to ISO 15004-2:2007) and 1.3% of MPE for the cornea and 1% of MPE for the retina (ANSI Z80.36-2016). These values reflect that the light beam is focused onto the cornea and widely spread on the retina. Note that up-to-date ISO and ANSI standards specify different MPE levels for corneal imaging at an 850 nm wavelength, which leads to the different safety margins. The subject's head was comfortably positioned with temple supports and a chin rest. While one eye was imaged, the second eye was fixating on a target.

**Image acquisition**

The operator choses the corneal layer to be imaged and extends the reference arm of interferometer to compensate for defocus, associated with this layer(*7*)·(*8*)·(*20*). Next the center of the interferometer FOV is aligned with the corneal apex using the joystick and controller on the vertical translation stage. The images are used as a feedback. More precisely, when imaging the anterior cornea, the operator looks at the mirror-like fringe patterns from the corneal surface. When imaging the posterior cornea, the operator looks at the mirror-like fringe pattern from the last corneal layer (endothelium). The high fringe density along the X or Y axes indicats the misalignment, which could be compensated for by shifting the interferometer along the corresponding direction until the fringe density decreases.

**Software**

Custom programs written in Labview 2014 were used for CF-OCT image acquisition and display. We utilized ImageJ(*21*) 1.51p for image display and measurements of cell density.

**Artifact suppression and contrast enhancement**

In order to remove the interference fringe artifacts from the endothelial images and reveal the endothelial mosaic in Figure 3, we collected a sequence of 40 camera images. After selecting 16 images with the highest signal level, we consequently subtracted them and obtained a 15-image stack. Then the stack was averaged, which reduced the sharpness of the fringes and, therefore, the spatial frequency in the image, associated with fringe borders. As a next step, the averaged image was transformed into the Fourier domain and the ring-shaped mask was applied to block the bright spots, corresponding to fringes, at the same time keeping the lowest frequencies to make the resulting image appear similar to that of specular microscopy (current clinical standard for endothelial cell counting). Inverting this Fourier image with suppressed artifacts revealed the cells.

Figure 4 was obtained without the image averaging step, however, the Fourier filter had to be extended to higher frequencies to suppress the non-averaged sharp interference fringes.

**Quantitative image analysis**

Manual endothelial cell counting in Figure 4 was done with the Multi-point Tool in ImageJ.

**Other instruments**

CF-OCT images were compared with clinical state-of-the-art instruments. The SNP was photographed using a clinical *in vivo* confocal microscope (IVCM; HRT II with Rostock cornea module; Heidelberg Engineering, GmbH, Germany) with 0.3 mm × 0.3 mm field of view (FOV). Prior to examination, one drop of a topical anesthetic, oxybuprocaine and one drop of a gel tear substitute, carbomer 0.3% (Gel-larmes, carbomer-974 P; Théa, Switzerland), with a refraction index similar to that of the cornea, were instilled in the eye. The endothelium was photographed using a clinical specular microscope (SM; SP-3000P, Topcon, Japan) with 0.25 x 0.5 mm FOV.

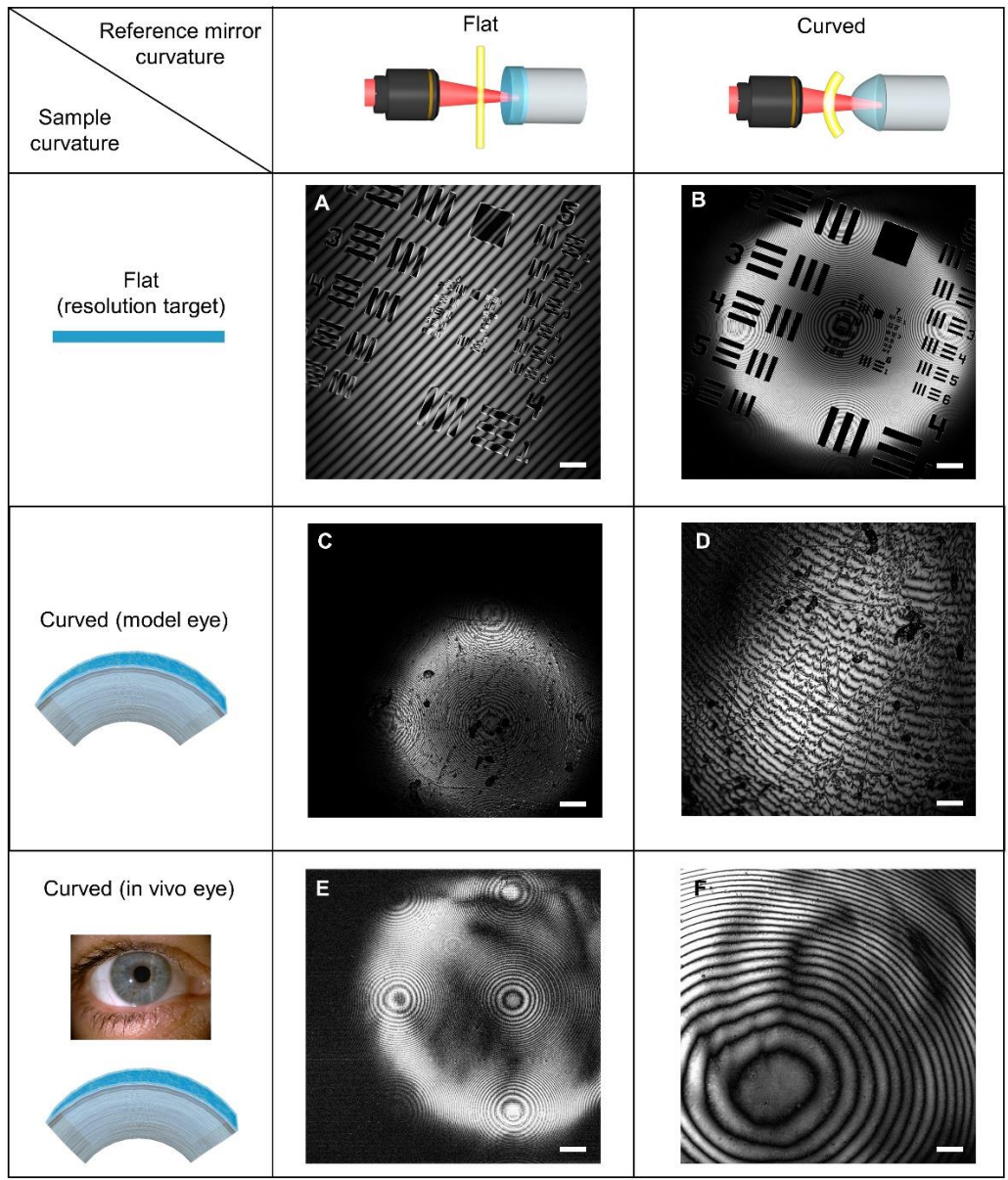

**Fig. S1. Correlation between interference fringe density and degree of curvature matching between the surfaces in the sample and reference arms of the interferometer.** The fringe density is small, when the curvatures of the reflecting surfaces in the sample and reference arms are similar, as is the case of the conventional TD-FF-OCT with identical flat reflectors (**A**) and CF-OCT with the curved reflector of 7.7 mm radius, matching to the shape of the artificial (**D**) or *in vivo* (**F**) anterior eye surfaces. Alternatively, the fringe density is high, when the curvatures of the reflecting surfaces in the sample and reference arms do not match, like in (**B**, **C**, **E**) cases. All scale bars are 0.1 mm.

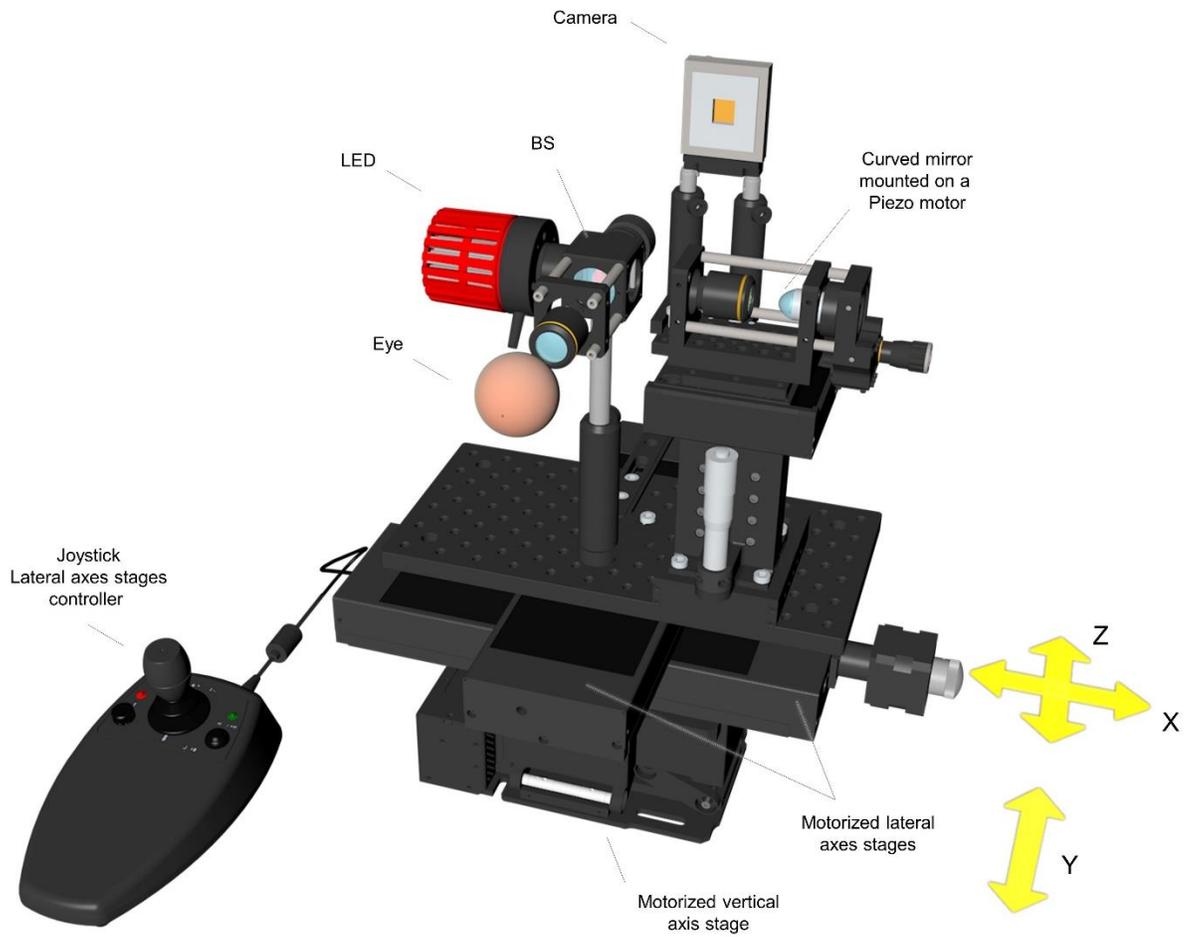

**Fig. S2. Curved-field OCT device.** The custom-made optical interferometer is mounted on three motorized stages. The stages, controlled with a joystick, are used to center the interferometer optics at the corneal apex, achieving a correct, curved optical sectioning. Small motorized stage beneath the reference arm is used to select the corneal layer to be imaged (e.g. SNP or endothelium).

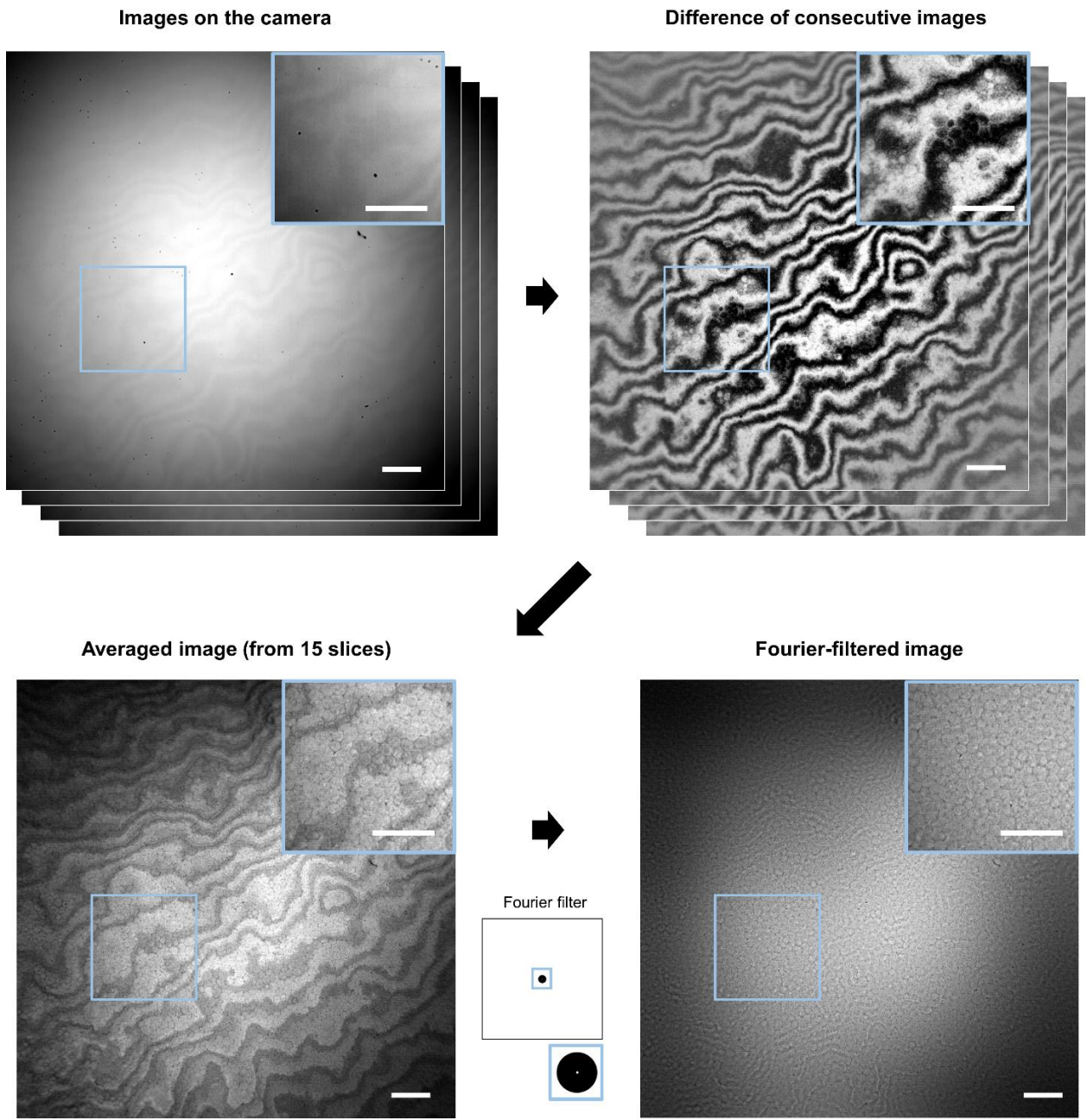

**Fig. S3. Retrieving endothelial cell mosaic from curved-field OCT camera images, obscured with interference fringes.** Interference fringes originating from the mirror-like reflection can be removed by subtracting the consecutive camera images and averaging, followed by filtering in the Fourier domain. All scale bars are 0.1 mm.

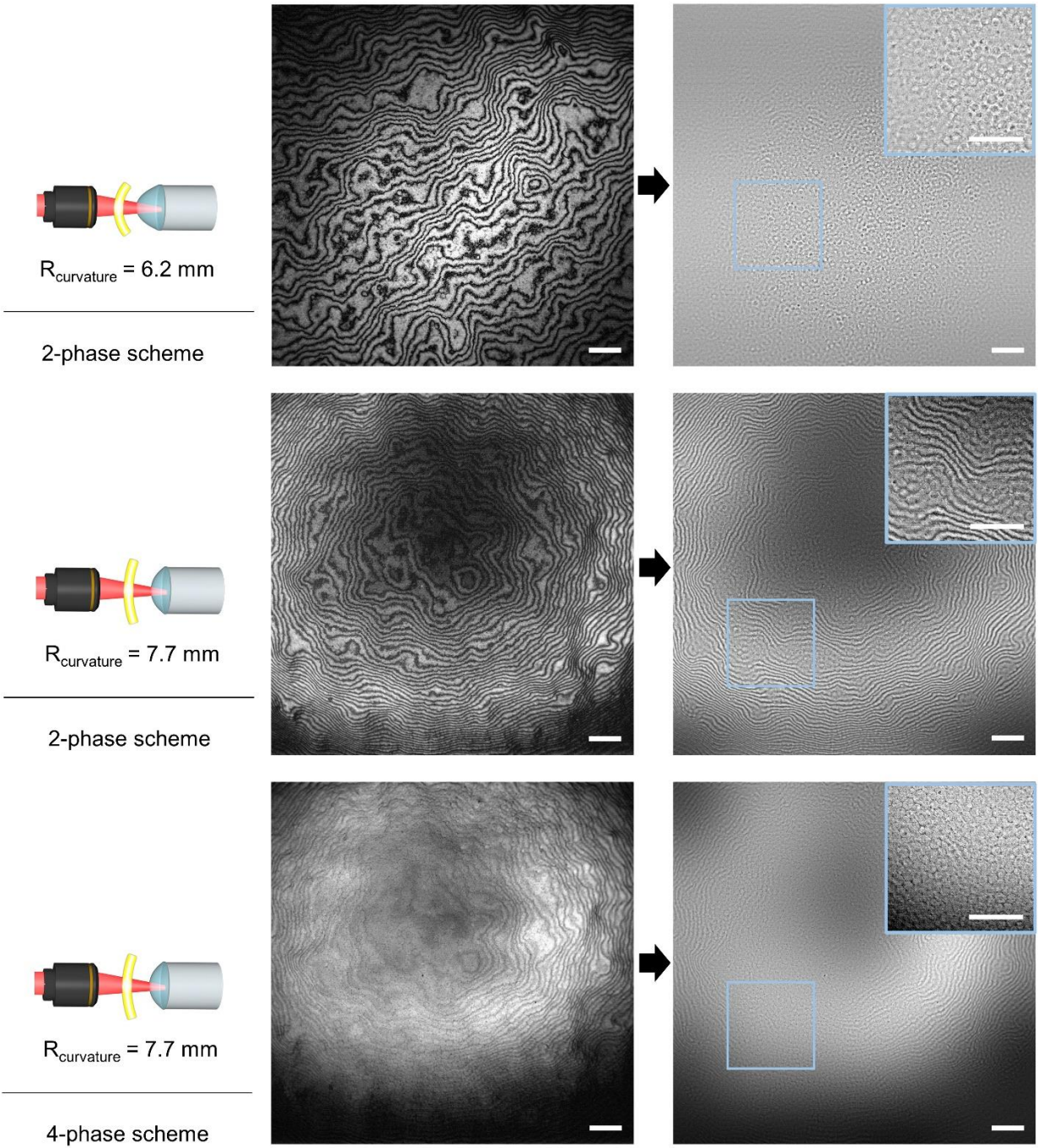

**Fig. S4. Comparison of endothelial images obtained with lenses of 6.2 mm (matching to the posterior corneal curvature) and 7.7 mm (matching to the anterior corneal curvature) radii of curvature.** The curvature mismatch is highlighted by the fringes with increased density at the border of the image, which are difficult to filter without affecting the underlying cells. Fringes can be still removed by either performing averaging before Fourier filtering or using a 4-phase tomographic image retrieval scheme. All scale bars are 0.1 mm.

19